\documentstyle[12pt,aaspp]{article}

\begin{document}

\newbox\grsign \setbox\grsign=\hbox{$>$} \newdimen\grdimen \grdimen=\ht\grsign
\newbox\simlessbox \newbox\simgreatbox
\setbox\simgreatbox=\hbox{\raise.5ex\hbox{$>$}\llap
     {\lower.5ex\hbox{$\sim$}}}\ht1=\grdimen\dp1=0pt
\setbox\simlessbox=\hbox{\raise.5ex\hbox{$<$}\llap
     {\lower.5ex\hbox{$\sim$}}}\ht2=\grdimen\dp2=0pt
\def\gtorder{\mathrel{\copy\simgreatbox}}
\def\ltorder{\mathrel{\copy\simlessbox}}
\def\simgreat{\mathrel{\copy\simgreatbox}}
\def\simless{\mathrel{\copy\simlessbox}}

\def\chaphead{}

\def\hut{Hubble type\ }
\def\vc{V$_{\rm C}$\ }
\def\mb{M$_{\rm B}$\ }
\def\av{A$_{\rm V}$\ }
\def\lamlam{$\lambda\lambda$}

\def\deg{$^\circ$}
\def\degrees{$^\circ$}
\def\Vlasov{collisionless Boltzmann\ }
\def\lsls{\ll}
\def\grgr{\gg}
\def\erf{\mathop{\rm erf}\nolimits} 
\def\eqv{\equiv}
\def\real{\Re e}
\def\imag{\Im m}
\def\ctrline#1{\centerline{#1}}
\def\spose#1{\hbox to 0pt{#1\hss}}
     
\def\={\overline}
\def\sections{\S}
\newcount\notenumber
\notenumber=1
\newcount\eqnumber
\eqnumber=1
\newcount\fignumber
\fignumber=1
\newbox\abstr
\newbox\figca     
\def\yyskip{\penalty-100\vskip6pt plus6pt minus4pt}
     
\def\numberpara{\yyskip\noindent}
     
\def\km{{\rm\,km}}
\def\kms{{\rm\ km\ s$^{-1}$}}
\def\kpc{{\rm\,kpc}}
\def\mpc{{\rm\,Mpc}}
\def\etal{{\it et al. }}
\def\eg{{\it e.g. }}
\def\ie{{\it i.e. }}
\def\cf{{\it cf. }}
\def\msun{{\rm\,M_\odot}}
\def\lsun{{\rm\,L_\odot}}
\def\rsun{{\rm\,R_\odot}}
\def\pc{{\rm\,pc}}
\def\cm{{\rm\,cm}}
\def\yr{{\rm\,yr}}
\def\au{{\rm\,AU}}
\def\AU{{\rm\,AU}}
\def\gm{{\rm\,g}}
\def\s{{\rmss}}
\def\dyne{{\rm\,dyne}}
     
\def\note#1{\footnote{$^{\the\notenumber}$}{#1}\global\advance\notenumber by 1}
\def\foot#1{\raise3pt\hbox{\eightrm \the\notenumber}
     \hfil\par\vskip3pt\hrule\vskip6pt
     \noindent\raise3pt\hbox{\eightrm \the\notenumber}
     #1\par\vskip6pt\hrule\vskip3pt\noindent\global\advance\notenumber by 1}
\def\propo{\propto}
\def\larrow{\leftarrow}
\def\rarrow{\rightarrow}
\def\sectionhead#1{\penalty-200\vskip24pt plus12pt minus6pt
        \centerline{\bbrm#1}\vskip6pt}
     
\def\Dt{\spose{\raise 1.5ex\hbox{\hskip3pt$\mathchar"201$}}}    
\def\dt{\spose{\raise 1.0ex\hbox{\hskip2pt$\mathchar"201$}}}    
\def\llangle{\langle\langle}
\def\rrangle{\rangle\rangle}
\def\ldotss{\ldots}
\def\del{\b\nabla}
     
\def\new{{\rm\chaphead\the\eqnumber}\global\advance\eqnumber by 1}
\def\ref#1{\advance\eqnumber by -#1 \chaphead\the\eqnumber
     \advance\eqnumber by #1 }
\def\last{\advance\eqnumber by -1 {\rm\chaphead\the\eqnumber}\advance
     \eqnumber by 1}
\def\eqnam#1{\xdef#1{\chaphead\the\eqnumber}}
     
\def\nfig{\chaphead\the\fignumber\global\advance\fignumber by 1}
\def\nfiga#1{\chaphead\the\fignumber{#1}\global\advance\fignumber by 1}
\def\rfig#1{\advance\fignumber by -#1 \chaphead\the\fignumber
     \advance\fignumber by #1}
\def\refindent{\par\noindent\parskip=3pt\hangindent=3pc\hangafter=1 }

\def\apj#1#2#3{\refindent#1,  {ApJ,\ }{\bf#2}, #3}
\def\apjsup#1#2#3{\refindent#1,  {ApJS,\ }{\bf#2}, #3}
\def\aasup#1#2#3{\refindent#1,  { A \& AS\ }{\bf#2}, #3}
\def\aas#1#2#3{\refindent#1,  { Bull. Am. Astr. Soc.,\ }{\bf#2}, #3}
\def\apjlett#1#2#3{\refindent#1,  { ApJL,\  }{\bf#2}, #3}
\def\mn#1#2#3{\refindent#1,  { MNRAS,\ }{\bf#2}, #3}
\def\mnras#1#2#3{\refindent#1,  { M.N.R.A.S., }{\bf#2}, #3}
\def\annrev#1#2#3{\refindent#1, { ARA \& A,\ }
{\bf2}, #3}
\def\aj#1#2#3{\refindent#1,  { AJ,\  }{\bf#2}, #3}
\def\phrev#1#2#3{\refindent#1, { Phys. Rev.,}{\bf#2}, #3}
\def\aa#1#2#3{\refindent#1,  { A \& A,\ }{\bf#2}, #3}
\def\nature#1#2#3{\refindent#1,  { Nature,\ }{\bf#2}, #3}
\def\icarus#1#2#3{\refindent#1,  { Icarus, }{\bf#2}, #3}
\def\pasp#1#2#3{\refindent#1,  { PASP,\ }{\bf#2}, #3}
\def\appopt#1#2#3{\refindent#1,  { App. Optics,\  }{\bf#2}, #3}
\def\spie#1#2#3{\refindent#1,  { Proc. of SPIE,\  }{\bf#2}, #3}
\def\opteng#1#2#3{\refindent#1,  { Opt. Eng.,\  }{\bf#2}, #3}
\def\refpaper#1#2#3#4{\refindent#1,  { #2 }{\bf#3}, #4}
\def\refbook#1{\refindent#1}
\def\science#1#2#3{\refindent#1, { Science, }{\bf#2}, #3}
     
\def\chapbegin#1#2{\eject\vskip36pt\par\noindent{\chapheadfont#1\hskip30pt
     #2}\vskip36pt}
\def\sectionbegin#1{\vskip30pt\par\noindent{\bf#1}\par\vskip15pt}
\def\subsectionbegin#1{\vskip20pt\par\noindent{\bf#1}\par\vskip12pt}
\def\topic#1{\vskip5pt\par\noindent{\topicfont#1}\ \ \ \ \ }
     
\def\ltsim{\mathrel{\spose{\lower 3pt\hbox{$\mathchar"218$}}
     \raise 2.0pt\hbox{$\mathchar"13C$}}}
\def\gtsim{\mathrel{\spose{\lower 3pt\hbox{$\mathchar"218$}}
     \raise 2.0pt\hbox{$\mathchar"13E$}}}
     
\def\sec{\hbox{$^s$\hskip-3pt .}}
\def\gg{\hbox{$>$\hskip-4pt $>$}}
\parskip=3pt
\def\gapprox{$_ >\atop{^\sim}$}     
\def\lapprox{$_ <\atop{^\sim}$}     
\def\apequal{\mathrel{\spose{\lower 1pt\hbox{$\mathchar"218$}}
     \raise 2.0pt\hbox{$\mathchar"218$}}}

\def\oforder{$\sim$} \def\inv{$^{-1}$}
\def\>={$\geq$} \def\<={$\leq$} \def\ks{km s\inv} \def\kms{km s\inv}
\def\lith{$h$} \def\sig{$\sigma$} \def\sigp{$\sigma^{\prime}_r$}
\def\meanz{$\overline \upsilon$} \def\nc{$N_c$} \def\rc{$r_c$}
\def\twidle{$\sim$} \def\sigmar{$\sigma_r$}
\def\Mstar{{M_R}^*}
\def\Mdot{M_{\odot}}

\title{THE ISOLATED ELLIPTICAL NGC 1132:  EVIDENCE FOR A MERGED
GROUP OF GALAXIES?}
\author{John S. Mulchaey}
\affil{The Observatories of the Carnegie Institution of Washington, 813 
Santa Barbara St., Pasadena, CA 91101}
\centerline{mulchaey@pegasus.ociw.edu}
\vskip 0.2cm

\centerline{and}
\vskip 0.2cm
\author{Ann I. Zabludoff}
\affil{UCO/Lick Observatory and Board of Astronomy and Astrophysics, 
University of California at Santa Cruz, Santa Cruz, CA 95064}
\centerline{aiz@ucolick.org}

\begin{abstract}
Numerical simulations predict that some poor groups of galaxies 
have merged by the present epoch into giant ellipticals (cf. Barnes 1989).
To identify the possible remnants of such mergers,
we have compiled a sample of nearby, isolated ellipticals 
(Colbert, Mulchaey, \& Zabludoff 1998).
ASCA observations of the first galaxy studied, NGC 1132, 
reveal an X-ray halo that extends out to at least $\approx$ 250 kpc h$_{\rm 100}^{-1}$.  
The temperature ($\sim$ 1 keV), metallicity ($\sim$ 0.25 solar) and 
luminosity ($\sim$ 2.5 $\times$ 10$^{42}$ h$_{\rm 100}^{-2}$ erg s$^{-1}$) of NGC 1132's X-ray halo
are comparable to those of poor group halos.
The total mass inferred from the X-ray emission,
$\sim$ 1.9$^{+0.8}_{-0.6}$ $\times$
10$^{13}$ h$_{\rm 100}^{-1}$ M$_{\odot}$, is also like that
of an X-ray detected group.  Optical imaging uncovers a dwarf galaxy
population clustered about NGC 1132 that is consistent in number
density and in projected radial distribution with that of an X-ray group.  
The similarities of NGC 1132 to poor groups in both the X-ray
band and at the faint end of the galaxy luminosity function, combined
with the 
deficit of luminous galaxies in the NGC 1132 field, are compatible with
the merged group picture.
Another possibility is that the NGC 1132 system is a `failed'
group ({\it i.e.}, a local overdensity in which other bright galaxies never
formed).

\end{abstract}


\section{Introduction}

The evolution of elliptical galaxies and of poor groups of galaxies is
probably linked.  Numerical simulations suggest that
a merged group will relax to form a single elliptical galaxy (Barnes 1989; 
Governato \etal 1991; Bode \etal 1993; Athanassoula, Makino, \& Bosma 1997).
Because the merger timescales for the brightest group members 
(M $\simless$ M$^*$) are a few tenths of a Hubble time (cf. Zabludoff \& Mulchaey 1998),
some groups have probably already
merged into elliptical remnants.  Observational clues to the
origin of these ellipticals may remain.
For example, the timescales for the cooling of the
group X-ray halo (cf. Ponman \& Bertram 1993; Ponman \etal 1994)
and the merger of dwarf group members are much
longer than for the merger of the brightest group galaxies. 
Therefore, a merged group could appear today as an
isolated elliptical with a group-like X-ray halo and a surrounding dwarf
galaxy population.

To identify merged group candidates, we have compiled a 
well-defined sample of nearby isolated elliptical galaxies (Colbert,
Mulchaey \& Zabludoff 1998).
Our sample was defined by examining the
environment of every galaxy classified as elliptical in the RC3 (de
Vaucouleurs \etal 1991) with z $<$ 0.03. We consider a galaxy \lq\lq
isolated\rq\rq \ when there is no other catalogued ({\it i.e.},
bright) galaxy within a projected radius of 1 h$_{\rm 100}^{-1}$
(h$_{\rm 100}$=H$_{\rm o}$/100) Mpc or a recessional velocity of $\pm
2000$ km s$^{-1}$.  Because surveys like the RC3 are notoriously
incomplete, we confirm the isolation of these galaxies by visual inspection
of the STScI Digitized Sky Survey plates and by comparisons with group
and cluster catalogs drawn from magnitude-limited redshift surveys
({\it e.g.}, CfA Redshift Survey, Huchra
\etal 1995).
Finally, we have obtained CCD images of the fields of these galaxies that 
confirm there are no other $\Mstar$ galaxies in the immediate vicinity
(typically within a $\sim$ 200 kpc h$_{\rm 100}^{-1}$ radius; Colbert et al. 1998). 
These criteria ensure that the environments of these
ellipticals are among the most rarefied ({\it i.e.}, have the lowest
bright galaxy densities) in the nearby universe.

In general, the galaxies in our isolated sample have not been
extensively studied. The majority of the galaxies do not have optical
imaging or spectroscopy available in the literature and none of them
have been studied in detail in the X-ray band. However, six of the
sample galaxies are in the ROSAT All-Sky Survey Bright Source Catalog
(Voges \etal 1996), indicating they are sufficiently bright for
future X-ray studies.  We have begun a program to study these objects
in detail with ASCA (Advanced Satellite for Cosmology and 
Astrophysics).  Here, we present the results of the
ASCA observations and optical imaging of our first target, NGC 1132, a
luminous (M$_{\rm R}$ $\approx$ -21.5 + 5 log h$_{\rm 100}$; q$_0$ = 0.5) elliptical 
with a recessional velocity of 6951 km s$^{-1}$ (RC3) and a 
stellar velocity dispersion of $\sigma = 253 \pm {14}$ km s$^{-1}$
(Tonry \& Davis 1981).

\section{X-ray Properties of NGC 1132}

NGC 1132 was observed by ASCA August 8-9, 1997. The data were screened
following the method described in Davis, Mulchaey \& Mushotzky (1998).
The resulting exposure times are 32.1 ksec for the 
GIS (Gas Imaging Spectrometer) and 34.6 ksec
for the SIS (Solid-State Imaging Spectrometer).
The GIS consists of two imaging gas scintillation proportional counters, each
with a 
circular field of view of diameter 50 arcminutes. The SIS
consists of two cameras each with four front-illuminated CCD chips.
However, because of the increase in the 
telemetry rates due to hot and flickering pixels, the four CCDs
can no longer be operated simultaneously. For our observation of NGC 1132,
a single CCD was used for each SIS camera.
The resulting field of view for the SIS cameras is approximately
11 arcminutes on a side.

A bright extended
X-ray source is clearly detected 
in both the SIS and GIS.
 The peak of the X-ray emission is
located within 30 arcseconds of the optical position of NGC 1132. This
offset is less than the 90\% error circle for ASCA, consistent with
the X-ray source being centered on NGC 1132 (Gotthelf 1996). In Figure 1, we
overlay the contours of the X-ray emission from the GIS2 on the STScI Digitized Sky
Survey image of the field.

To determine the extent of the X-ray emission, we construct an
azimuthally-averaged surface brightness profile from the GIS2 data
(Figure 2). Beyond a radius of $\sim$ 14$'$ the profile reaches a 
constant flux of $\sim$ 1.6 $\times$ 10$^{-4}$ counts s$^{-1}$ arcmin$^{-2}$.
We adopt this flux as the background level.
The surface brightness profile 
reaches a level of 20\% of the background at a radius of 
$\sim$ 12$'$ (corresponding to 243 h$_{\rm 100}^{-1}$ kpc). As this 
level is approximately the systematic error in the background, we 
adopt this radius as the maximum extent of the diffuse gas. 
We note that the physical extent of the hot gas in NGC 1132 is comparable to that
of the X-ray halos of poor groups of galaxies
({\it e.g.}, Mulchaey \etal 1996; Ponman \etal 1996). 

We then use the XPROF program (Ptak 1997), which properly accounts
for the GIS point spread function, to fit the profile with a standard
hydrostatic-isothermal beta model (Cavaliere \& Fusco-Femiano 1976) :

\centerline{S(r)=S$_{\rm o}$
(1.0 + (r/r$_{\rm core}$)$^2$)$^{-3\beta + 0.5}$}

\noindent{where the free parameters are S$_{\rm o}$ (the central surface
brightness), r$_{\rm core}$ (the core radius), and $\beta$. We fix the background level
in the fit to the value given above. The best fit to the profile gives
$\beta$=0.83$^{+0.21}_{-0.13}$ and r$_{\rm core}$=4.65$'$$^{+1.42}_{-0.99}$
(94$^{+28.8}_{-20.0}$ h$_{\rm 100}^{-1}$ kpc). Although the $\beta$ model fit is
formally acceptable ($\chi$$^{2}_{\rm R}$ $=$1.4), 
Figure 2 indicates that the model does not reproduce the flux levels at all 
radii and, in particular, it underestimates the central emission. 
This discrepancy is probably an indication that more complex models are required to 
describe the density profile.
We note that the
surface brightness profiles of groups derived from higher spatial resolution ROSAT PSPC
observations suggest the presence of two X-ray components in these
systems (Mulchaey \& Zabludoff 1998). 
However, given the poor spatial
resolution of the ASCA data, we do not attempt to fit more complicated
models here.
Future high spatial resolution
images with AXAF (Advanced X-ray Astrophysics Facility) should provide 
better constraints on the gas density profile in NGC 1132 and allow us to search for the 
presence of multiple components in this system.

To determine the temperature and metallicity of the diffuse X-ray
emission we extracted SIS and GIS spectra.  As the SIS detectors were 
operated in 1-CCD mode, only the inner regions of the diffuse gas were
covered. 
Thus, we are able to extract SIS spectra only out to a radius of 4$'$
(81 h$_{\rm 100}^{-1}$ kpc).  For the GIS detectors, we extracted two
spectra, one over the region covered by the SIS 
({\it i.e.,} 4$'=
81$ h$_{\rm 100}^{-1}$ kpc) and a
second out to a radius of 12$'$ (243 h$_{\rm 100}^{-1}$ kpc),
corresponding to the maximum measured extent of the X-ray emission.
The spectra were fit with a Raymond-Smith model (Raymond \& Smith 1977) with the gas
temperature and metal abundance as free parameters. The absorbing
column was fixed at the Galactic value of N$_{\rm H}$ $=$ 5.2 $\times$
10$^{20}$ cm$^{-2}$ (Stark \etal 1992). A simultaneous fit to the
SIS and GIS (radius=4$'$) data gives a temperature of
T=1.11$\pm{0.02}$ keV and a metal abundance of 0.58$^{+0.17}_{-0.12}$
solar. A fit to the GIS data out to radius=12$'$ gives a slightly lower
best-fit temperature and metal abundance (T=1.04$^{+0.08}_{-0.12}$
keV, abundance=0.25$^{+0.25}_{-0.14}$ solar), but these values are
still consistent within the errors with the spectral fit for the inner
region. 
Based on the GIS data, the bolometric X-ray luminosity of NGC 1132
out to a radius of 12$'$ (243 h$_{\rm 100}^{-1}$ kpc)
is $\sim$ 2.5 $\times$ 10$^{42}$ h$_{\rm 100}^{-2}$ erg s$^{-1}$.
Overall, the temperature, metallicity and luminosity of the diffuse X-ray emission in 
NGC 1132 are similar
to those of the intragroup medium in X-ray detected poor groups as determined by
ASCA observations (Davis et al. 1998; Fukazawa et al. 1998).

The X-ray data can also be used to estimate the total mass of the NGC
1132 system.  Assuming that the hot gas is in hydrostatic equilibrium and
is isothermal, the total mass of the group at radius r	is given by (e.g. Fabricant, Rybicki
\& Gornestein 1984):
 
\centerline{M$_{\rm total}$ ($<$r) = --${kT_{\rm gas}(\rm r)}\over{G{\mu}m_p}$
[${{dlog{\rho}}\over{dlog{\rm r}}} + {{dlogT}\over{dlog{\rm r}}}$] r}
 
\vskip 0.5cm

\noindent{where $T_{\rm gas}$(r) is the gas temperature at radius r, 
$\rho$ is the gas density, k is Boltzmann's constant, G is the gravitational
constant, $\mu$ is the mean molecular weight, and $m_p$ is the mass of the proton.
Adopting a $\beta$ model for the gas density profile and assuming that the 
gas is isothermal ({\it i.e.,} ${dlogT}\over{dlog{\rm r}}$ = 0), this equation reduces to:

\centerline{M$_{\rm total}$ ($<$r) = ${3kT_{\rm gas}\beta {\rm r_{\rm core}}}\over{G{\mu}m_p}$
{${x^3}\over{1+x^2}$}} 

\noindent{where $x=$r/r$_{\rm core}$.
Applying this equation to NGC 1132 and assuming the 
density profile derived from the $\beta$ model fit to the surface brightness profile, 
the total mass out to a radius of
12$'$ (243
kpc) is $\approx$ 1.9$^{+0.8}_{-0.6}$ $\times$ 10$^{13}$ h$_{\rm 100}^{-1}$
M$_{\odot}$. 
This mass is approximately the average mass found for poor
groups observed by ROSAT (Mulchaey \etal 1996) and approximately 10 times the mass
of the galaxy estimated from its internal velocity dispersion and 
effective radius (Tonry \& Davis 1981).}

\section{Environment of NGC 1132}
In the last section, we showed that NGC 1132 has a group-like X-ray halo, 
as might be expected if this galaxy is the remnant of a merged group.
The number counts and distribution of galaxies in the NGC 1132 field
provide additional tests of the merged group hypothesis.
In this section, we compare
the projected radial profile of faint galaxies in the NGC 1132
field with that of a composite of
five, nearby X-ray detected groups to determine if NGC 1132 lies in a group-like
system of dwarfs.
These comparison groups are all of the groups
in the Zabludoff \& Mulchaey (1998) sample that have X-ray
luminosities similar to NGC 1132 and for which we have optical imaging data. The
five groups are NGC 2563, NGC 4325, NGC 5129, HCG 42 and HCG 62.

Images were obtained for NGC 1132 and the comparison X-ray groups in
October 1996 at the 40-inch telescope at Las Campanas Observatory. The total
exposure time for each field was 5 minutes using a Tek 2048 CCD chip
and a Harris R filter. The images were reduced using standard
techniques in IRAF and flux calibrated with standard star
observations.  For each field, the program SExtractor (Bertin \&
Arnouts 1996) was used to classify objects as stars or galaxies and to
measure total magnitudes.  Total magnitudes were measured using a
method proposed by Kron (1980).  Examination of plots of total
magnitude versus total surface area indicate that SExtractor can
cleanly separate galaxies from stars down to apparent R magnitudes of
$m_R \approx 19.5$. This magnitude limit corresponds to an absolute magnitude of
$\approx -15 + 5$ log h$_{\rm 100}$ at the distance of 
our most distant object (NGC
4325) and we adopt this value as the limit of our analysis.
For the purposes of this analysis, we consider only galaxies projected within
178 h$_{\rm 100}^{-1}$ kpc 
of the field centers (equivalent to the 
radius=9.14$'$ field-of-view of our NGC 1132 image).

In Figure 3 we plot the projected surface density of faint 
galaxies with apparent R magnitudes 
between 17.22 and 19.22 in the NGC 1132 field (open circles).
If these galaxies all lie at the distance of NGC 1132, this magnitude range
corresponds to $-15 + 5$ log h$_{\rm 100}$
$<$ M$_{R}$ $<$ $-17 + 5$ log h$_{\rm 100}$. 
The Figure shows that the surface density of faint galaxies increases
within the central 60 h$^{-1}_{\rm 100}$ kpc, reaching a 
peak density of $\sim 0.0032 \pm{0.0015}$ h$^{-2}$ kpc$^{-2}$.
The clustering of faint galaxies around NGC 1132 suggests that these
galaxies are true dwarfs and are physically associated with the giant elliptical.

By assuming that the faint galaxies
in each group field are at the distance of the group, we construct a composite group 
radial profile for galaxies with 
$-15 + 5$ log h$_{\rm 100}$ $<$ M$_{R}$ $<$ $-17 + 5$ log h$_{\rm 100}$ (Figure 3,
filled circles).
This assumption is reasonable because, for both NGC 1132
and the composite group, the background level estimated from
the fall-off in the profile at radii $\simgreat 100$ h$_{\rm 100}^{-1}$ 
kpc is small compared to the peak \footnote{Counting statistics give $\sim 3.0\sigma$
as the significance of the peak above the background in the NGC 1132 field.  The fact that the 
dwarfs are clustered within the inner 60h$^{-1}_{\rm 100}$ kpc 
of NGC 1132 suggests that the true significance of the peak is actually much higher.}.
Figure 3 demonstrates that {\it both} 
the number density and spatial distribution of
dwarfs in the NGC 1132 field are comparable to those of the X-ray groups.

\section{Discussion}

The X-ray and optical observations discussed in the last two sections
provide important insight into the NGC 1132 system. The X-ray
properties of NGC 1132, such as gas extent, temperature, metallicity,
and X-ray luminosity are all indistinguishable from those of X-ray
detected groups. Furthermore, the total mass of the group as inferred
from the X-ray halo is typical of poor groups ({\it e.g.}, Mulchaey \etal
1996). From an X-ray perspective, NGC 1132 would be considered a
group.  Although the faint galaxies in the vicinity of NGC 1132 are
clustered and their counts are group-like, there are no other bright 
($M_R < {M_R}^*$) galaxies in
the field.  Thus, NGC 1132's optical properties would not qualify it as a poor group.

The X-ray and optical observations might be reconciled if NGC 1132 is
the product of a merged group.  If NGC 1132 formed from the merger of the brightest 
group members, then its internal velocity dispersion is probably
like that of the original group ($\approx 250$ km s$^{-1}$).
For a group with this velocity dispersion (cf. Zabludoff \& Mulchaey 1998),
the timescale for 
a M$^{*}$ galaxy to merge with another galaxy is roughly a few tenths of a Hubble time
t$_{\rm H}$.
The dynamical friction timescale for a M$^{*}$ galaxy to fall into the center from an
initial radius of $\sim 200$ h$_{\rm 100}^{-1}$ kpc 
(the extent of the observed NGC 1132 field) is similar.
Thus, the luminous galaxies of a poor group
could have merged into a central giant elliptical by the 
present time.  

The dynamical evolution of fainter galaxies is slower.  For example,
the galaxy-galaxy merger and dynamical friction timescales
for a M$^{*}_{R} + 2$ ($\approx -18.5 + 5$log h$_{\rm 100}$)
galaxy in the same $\approx 250$ km s$^{-1}$ group
are $\simgreat$ t$_{\rm H}$ (Zabludoff \& Mulchaey 1998).  Because the timescales for
even fainter galaxies are longer still, we would expect many dwarfs
({\it i.e.}, $-17 + 5$log h$_{\rm 100} <$ M$_{\rm R}$ $< -15 + 5$log h$_{\rm 100}$)
in the NGC 1132 system to have survived until the present day.
Therefore, the group-like, clustered dwarf population 
and the deficit of luminous galaxies 
within $\sim 200$ h$_{\rm 100}^{-1}$ kpc of NGC 1132 
are consistent with the merged group hypothesis.

A useful estimate of the lifetime of NGC 1132's X-ray halo is the 
cooling time, t$_{\rm cool}$ $\equiv$ (d ln T$_{\rm gas}$/dt)$^{-1}$, where 
T$_{\rm gas}$ is the temperature of the gas.
At X-ray emitting temperatures, t$_{\rm cool}$ $\propto$ T$_{\rm gas}$
 n$^{-1}$ $\Lambda$(T$_{\rm gas}$)$^{-1}$, where n is the gas number
density and $\Lambda$(T$_{\rm gas}$) is the cooling function.
Given our best fit $\beta$ model for the surface brightness 
profile, the gas density at 200 h$_{\rm 100}^{-1}$ 
kpc from the 
center of NGC 1132 is $\sim$ 3 $\times$ 10$^{-4}$ cm$^{-3}$. The
temperature of the extended gas is $\sim$ 10$^7$ K. Using the cooling 
function in Raymond, Cox \& Smith (1976) for a 
10$^7$ K gas, we estimate a cooling time for the halo gas of
roughly a Hubble time.
Therefore, while the bright galaxies may have had enough time to
merge into a single object, the large-scale X-ray halo remains basically unchanged from the 
time when the system was a group. 
Both the dwarf-to-giant ratio and the
X-ray properties of NGC 1132 support the picture in which this giant elliptical
was once a poor group.

On the other hand, it is possible that the NGC 1132 system
{\it formed} with a deficit of luminous galaxies. In this \lq\lq failed
group\rq\rq \ model, the majority of the stellar baryons were initially used
up in a single luminous galaxy, rather than in several M$^{*}_{R}$
objects.  Hattori \etal (1997) have recently reported the discovery
of another massive system that seems to lack galaxies. Based on ASCA
observations, AXJ2019+1127 appears to be a luminous X-ray cluster at
redshift z $\sim$ 1.  However, optical imaging of the field around the
X-ray source reveals only one luminous galaxy. The implied
mass-to-light ratio of AXJ2019+1127 is M/L$_{\rm B}$ $\approx$ 6,600
h$_{\rm 100}$.  What distinguishes this system from NGC 1132 is that
AXJ2019+1127 is lacking in both the number of galaxies and in total
stellar light, while the mass-to-light ratio of NGC 1132 (M/L$_{\rm
R}$ $\sim$ 300 h$_{\rm 100}$) is comparable to poor groups of similar
mass (the 
average M/L$_{\rm R}$ for the five X-ray groups discussed
in Section 3 is $\sim$ 340
h$_{\rm 100}$). Therefore, the NGC 1132 system is not \lq\lq
dark\rq\rq \ compared to groups, but rather has a high proportion of
its stellar mass in a single object.

Regardless of the origin of isolated ellipticals, further studies of
these galaxies are clearly warranted. Because the dark halos of giant
ellipticals in denser environments like clusters are likely to be
tidally-truncated by the cluster potential, isolated ellipticals may
provide our only means of studying the halos of bright ellipticals
before truncation. In the case of NGC 1132, the population of dwarf
galaxies can be used as test particles to study the dynamics of the
extended dark halo. 
Further X-ray observations of NGC 1132 and other isolated
ellipticals should also prove fruitful. For example, AXAF 
observations
should reveal whether the X-ray surface brightness profiles in these
galaxies show the two-component structure characteristic of X-ray
detected groups (Mulchaey \& Zabludoff 1998). 

\acknowledgments

We thank David Davis and Andy Ptak for their assistance in fitting the
X-ray surface brightness profile and Dennis Zaritsky for detailed 
comments on the text. We also acknowledge useful
discussions with James Colbert, Ron Marzke, Richard Mushotzky,
Gus Oemler and
Scott Trager. We want to especially thank the anonymous referee for a
a very careful report that significantly improved this
paper. 
This research involved the use of the HEASARC and NED databases.  J.
S. M. acknowledges partial support for this program from NASA grant
NAG 5-3529 and from a Carnegie postdoctoral fellowship.  AIZ
acknowledges support from the Carnegie and Dudley Observatories, the
AAS, NSF grant AST-95-29259, and NASA grant HF-01087.01-96A.

\vfill\eject
\begin{figure}
\epsfysize=5in
\centerline{\epsfbox{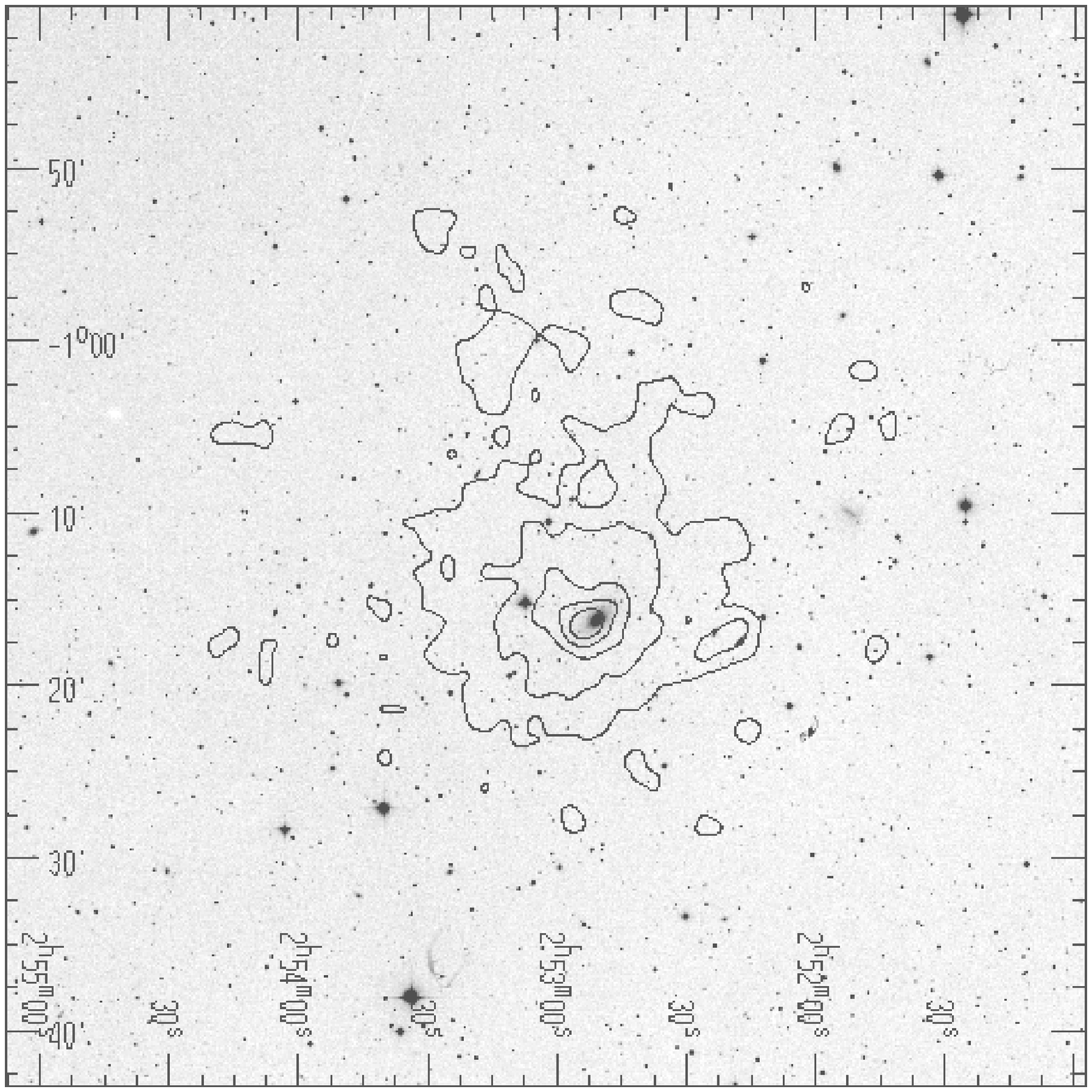}}
\caption{Contour map of the X-ray emission (ASCA GIS2) in NGC 1132 overlaid on the STScI 
Digital Sky Survey (field of view 1$^{\rm o}$ $\times$ 1$^{\rm o}$).
The coordinate scale is in J2000. The contours correspond to 5, 15,
30, 50 and 60 $\sigma$ above the background level. The ASCA data have
been smoothed with a Gaussian profile of width=30$''$. While the
center of the X-ray emission is slightly offset from the optical
center of NGC 1132, the observed offset is less than the 90\% pointing
accuracy of ASCA, so the X-ray center is consistent with the optical
position.}
\end{figure}

\vfill\eject
\begin{figure}
\epsfysize=5in
\centerline{\epsfbox{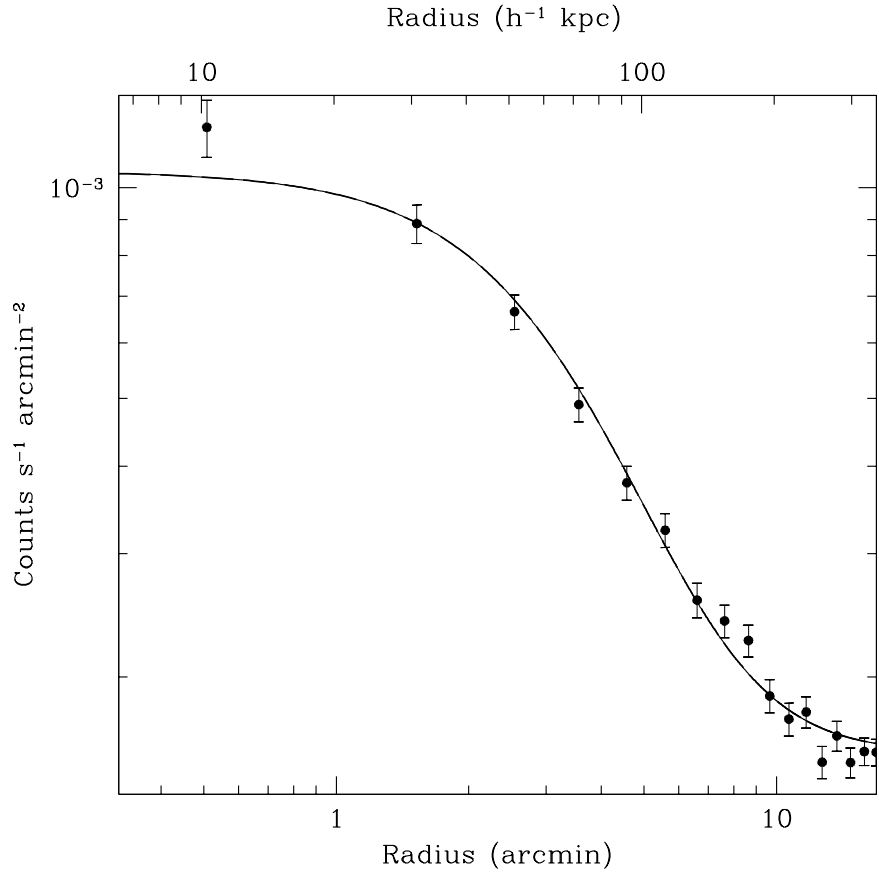}}
\caption{Surface brightness profile for the X-ray emission in NGC 1132 
derived from the ASCA GIS2 data. The solid line shows the best fit $\beta$ model.} 
\end{figure}

\vfill\eject

\begin{figure}
\epsfysize=5in
\centerline{\epsfbox{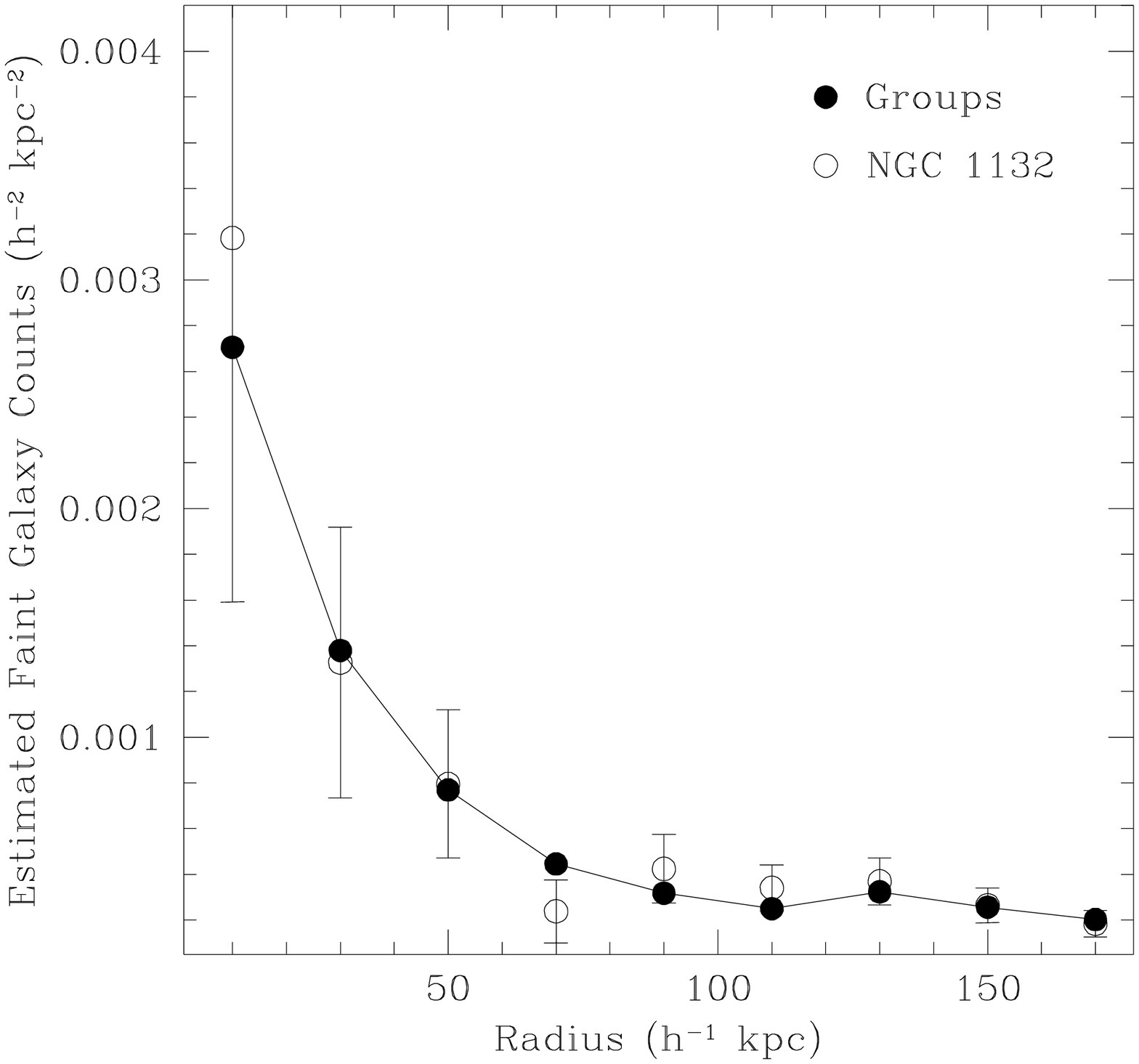}}
\caption{The radial surface density profile 
of faint galaxies in the NGC 1132 field (open circles) compared with the composite profile of
five X-ray groups (filled circles).  We include
only galaxies with estimated absolute magnitudes between 
-17 + 5 log h$_{\rm 100}$ and -15 + 5 log h$_{\rm 100}$.
At large radii ({\it i.e.}, $>$ 100 $h_{\rm 100}^{-1}$ kpc), the estimated galaxy counts in
both NGC 1132 and the composite X-ray group are consistent with the rather uncertain
background counts expected 
from R-band galaxy-count surveys (Koo \& Kron 1992). However, 
there is a statistically significant excess of galaxies in this magnitude range near the
centers of both NGC 1132 and the composite group. This excess suggests that many of the 
faint galaxies within 60 $h_{\rm 100}^{-1}$ kpc of the center of NGC 1132 are dwarfs
physically associated with NGC 1132. The error bars plotted for the NGC 1132 data
are based on the Poisson counting statistics. The error bars for the composite group
data are considerably smaller (on average by a factor of $\sim$ 2.6) and are not plotted.}  
\end{figure}


\begin{references}
\reference{} Athanssoula, E., Makino, J., \& Bosma, A. 1997, MNRAS, 286, 825
\reference{} Barnes, J. E. 1989, Nature, 338, 123
\reference{} Bertin, E., \& Arnouts, S 1996, AA Suppl., 313, 21
\reference{} Bode, P. W., Cohn, H. N., \& Lugger, P. M. 1993, ApJ, 416, 17
\reference{} Cavaliere, A., \& Fusco-Femiano, R. 1976, AA, 49, 137
\reference{} Colbert, J. W., Mulchaey, J. S., \& Zabludoff, A. I. 1998,
in preparation
\reference{} Davis, D. S., Mulchaey, J. S., \& Mushotzky, R. F. 1998, ApJ, in press.
\reference {} de Vaucouleurs, G., de Vaucouleurs, A., Corwin, H. G. Jr., Buta, R. J.,
Paturel, G., \& Foqu\'e, P. 1991, {\it Third Reference Catalogue of Bright Galaxies} 
(New York: Springer-Verlag)(RC3)
\reference{} Governato, F., Bhatia, R., \& Chincarini, G. 1991, ApJ, 371, 15
\reference{} Gotthelf, E. 1996, in {\it ASCANews}, No. 4
\reference{} Fabricant, D., Rybicki, G. B., \& Gorenstein, P. 1984, ApJ, 286, 186
\reference{} Fukazawa, Y. et al. 1998, PASJ, 50, 187
\reference{} Hattori, M. et al. 1997, Nature, 388, 146
\reference{} Huchra, J. P., Geller, M. J., \& Corwin, H. G. 1995, ApJ Suppl, 99, 391
\reference{} Koo, D. C., \& Kron, R. G. 1992, ARAA, 30, 613
\reference{} Kron, R. G. 1980,  ApJ Suppl, 43, 305
\reference{} Mulchaey, J. S., Davis, D. S., Mushotzky, R. F., \& Burstein, D. 1996, ApJ, 456, 80
\reference{} Mulchaey, J. S., \& Zabludoff, A. I. 1998, ApJ, 496, 73
\reference{} Ponman, T. J., Allan, D. J., Jones, L. R., Merrifield, M., McHardy, I. M.,
Lehto, H. J., \& Luppino, G. A. 1994, Nature, 369, 462
\reference{} Ponman, T. J., Bourner, P. D. J., Ebeling, H., \& Bohringer, H. 1996,
MNRAS, 283, 690
\reference{} Ponman, T. J., \& Bertram, D. 1993, Nature, 363, 51
\reference{} Ptak, A. 1997, PhD thesis, University of Maryland, College Park
\reference{} Raymond, J. C., Cox, D. P., \& Smith, B. W. 1976, ApJ, 204, 290
\reference{} Raymond, J. C., \& Smith, B. W. 1977, ApJ Suppl., 35, 419
\reference{} Stark, A. A., et al. 1992, ApJ Suppl, 79, 77
\reference{} Tonry, J. L., \& Davis, M. 1981, ApJ, 246, 666
\reference{} Voges, W. et al. 1996, IAU Circular, 6420, 2
\reference{} Zabludoff, A. I., \& Mulchaey, J. S. 1998, ApJ, 496, 39
\end{references}
\end{document}